\documentclass{PoS}
\newcommand{\scale}{0.275}

\title{The Influence of Instantons on the Quark Propagator }

\ShortTitle{The Influence of Instantons on the Quark Propagator }

\author{\speaker{Daniel Trewartha}\\
        CSSM, School of Chemistry and Physics, University of Adelaide, SA, 5005, Australia\\
        E-mail: \email{daniel.trewartha@student.adelaide.edu.au}}
\author{Waseem Kamleh\\
		CSSM, School of Chemistry and Physics, University of Adelaide, SA, 5005, Australia}
\author{Derek Leinweber\\
        CSSM, School of Chemistry and Physics, University of Adelaide, SA, 5005, Australia}
\author{Peter Moran}

\abstract{We use over-improved stout-link smearing to investigate the presence and nature of instantons on the lattice. We find that smearing can remove short-range effects with little damage to the long-range structure of the gauge field, and that after around 50 sweeps this process is complete. There are more significant risks for very high levels of smearing beyond 100 sweeps. We are thus able to produce gauge configurations dominated by instanton effects. We then calculate the overlap quark propagator on these configurations, and thus the non-perturbative mass function. We find that smeared configurations reproduce the majority of dynamical mass generation, and conclude that instantons are primarily responsible for the dynamical generation of mass.}

\FullConference{The 30th International Symposium on Lattice Field Theory\\
                 June 24 -- 29,  2012\\
                 Cairns, Australia}

\begin{document}
\section{Introduction}
Instantons have long been believed to be an essential component of the long-distance physics of the QCD vacuum, and lattice QCD provides an opportunity to gain unique insight into their role. Isolating their effects requires some care however; a UV filter is required in order to dampen short-range fluctuations, examples of which historically have included cooling \cite{Berg:1981nw,Teper:1985rb,Ilgenfritz:1985dz}, APE smearing \cite{Albanese:1987ds}, HYP smearing \cite{Hasenfratz:2001hp} and stout link smearing \cite{Morningstar:2003gk}. These algorithms, however, can destroy the instanton content of the vacuum, and so we will use over-improved stout-link smearing \cite{Moran:2008ra,Garcia Perez:1993ki}, tuned to preserve instantons. This is briefly described in section \ref{sec:Smearing}. We will then calculate the overlap quark propagator on these configurations in section \ref{sec:Results}, and compare the non-perturbative mass function on smeared and unsmeared configurations.
\newline
\section{Smearing}
\label{sec:Smearing}
In order to isolate instanton effects, we wish to remove short-range noise from the lattice, leaving just the underlying long-range structure of the gluon field. We do this by smearing; essentially averaging gauge links with their neighbours. Since an instanton ensemble is an approximate classical solution to the field equations, instantons should remain, while deviations from the classical solution are removed. Unfortunately, most smearing algorithms shrink instantons on the lattice, and eventually destroy them. We will thus use over-improved stout link smearing \cite{Garcia Perez:1993ki}, which is tuned to preserve instantons of size greater than $a$. This necessarily introduces the problem of instantons being enlarged by the smearing process, and so we must be careful that we use an amount of smearing small enough so as not to excessively distort the configurations. Applying successive sweeps of smearing, we should create configurations dominated by instanton degrees of freedom. \newline
We investigate the instanton content of configurations by finding local maxima of the action, then fitting the instanton action density \cite{Atiyah:1978physlett},
\begin{equation}
S_{0}(x) = \xi \frac{6}{\pi^{2}} \frac{\rho^{4}}{((x-x_{0})^{2}+\rho^{2})^{4}},
\end{equation}
 around these points, with $\rho$ the instanton radius. The topologically non-trivial objects observed on the lattice are expected to have a higher action than the continuum, and so we introduce the parameter $\xi$, allowing the shape of the action, rather than the height, to determine the fit.
This gives us a list of instanton candidates on the lattice, which we can then compare to the theoretical relationship between an instanton's radius and topological charge at the centre,
\begin{equation}
q(x_{0})=Q\frac{6}{\pi^{2}\rho^{4}},
\end{equation}
with $Q = \mp 1$ for an (anti)instanton.
\begin{figure}
\label{fig:radiusvcharge}
\includegraphics[scale=\scale,angle=90]{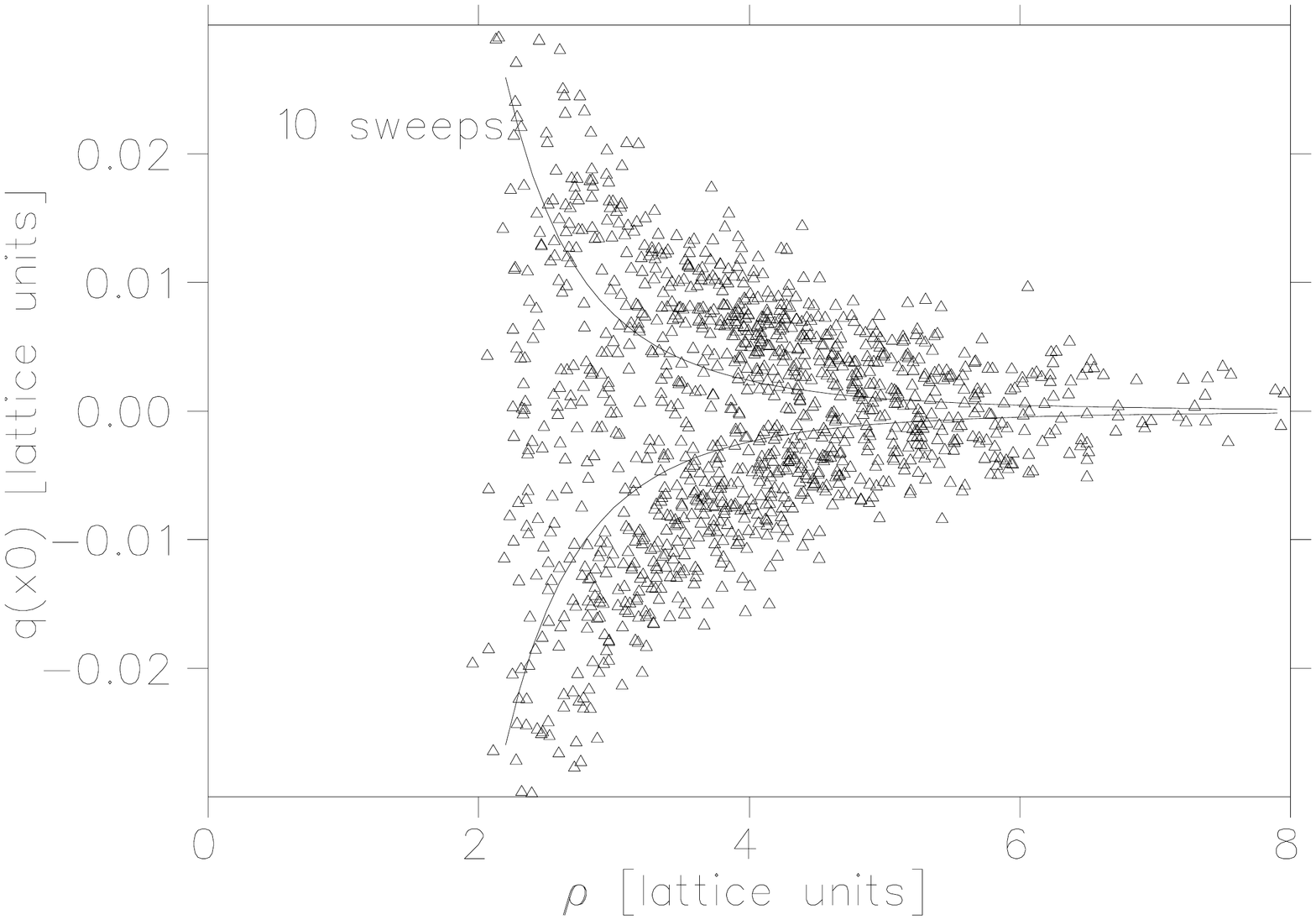}
\includegraphics[scale=\scale,angle=90]{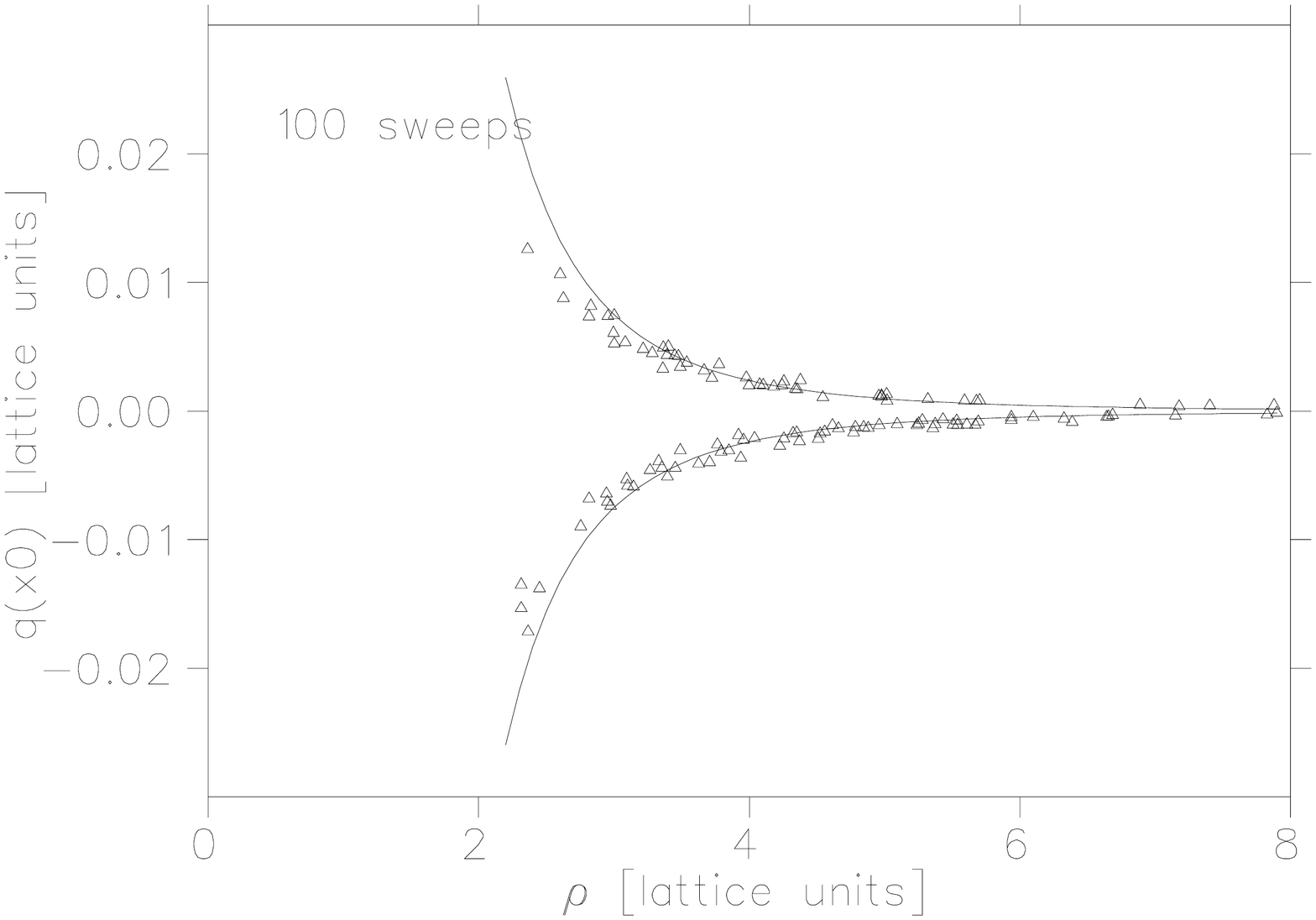}
\caption{Instanton candidate fitted radius vs topological charge at centre after 10 (left) and 100 (right) sweeps of over-improved stout-link smearing.}
\end{figure}
\par
\begin{figure}
\includegraphics[scale=\scale,angle=90]{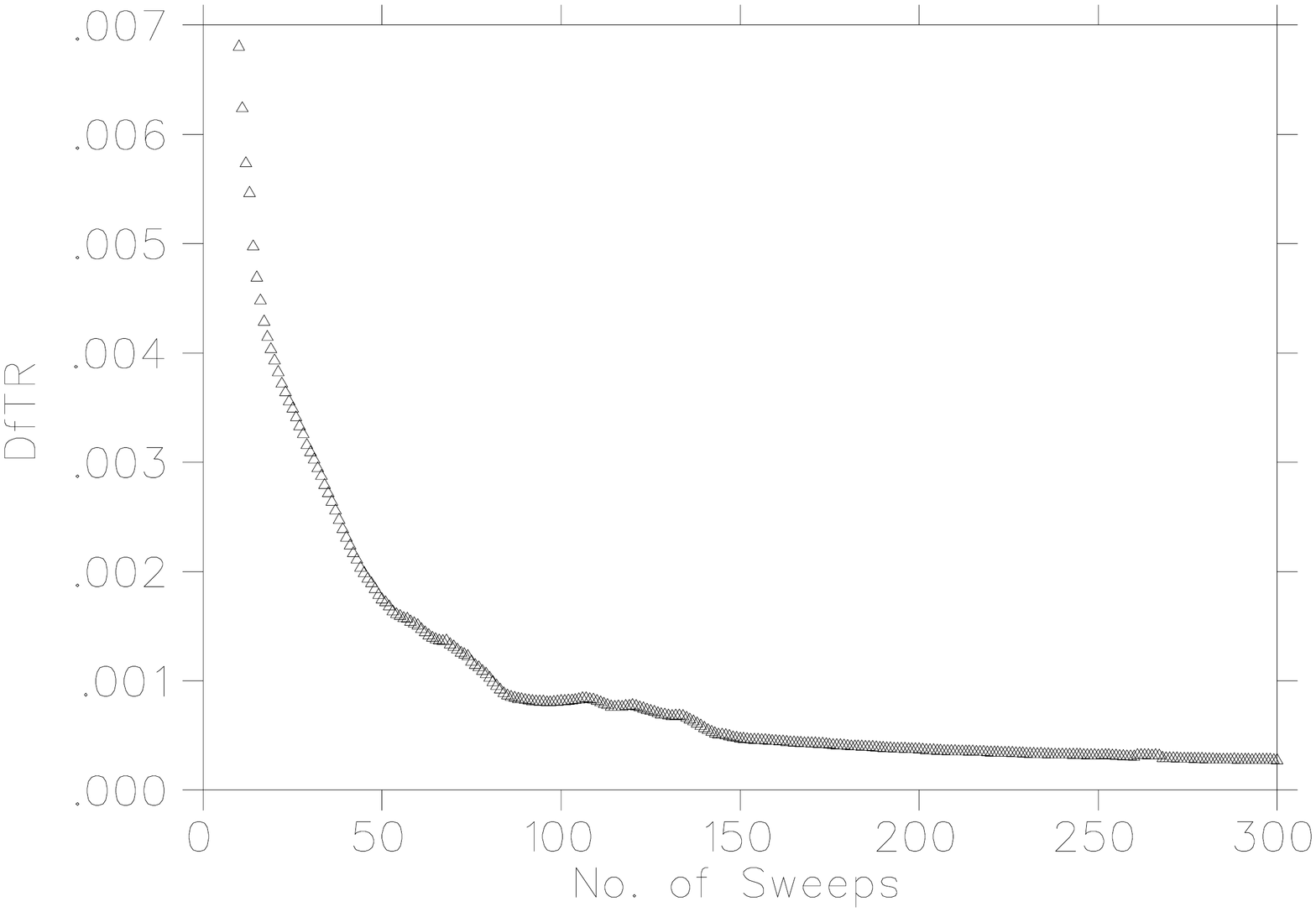}
\label{fig:tensweeps}
\includegraphics[scale=\scale,angle=90]{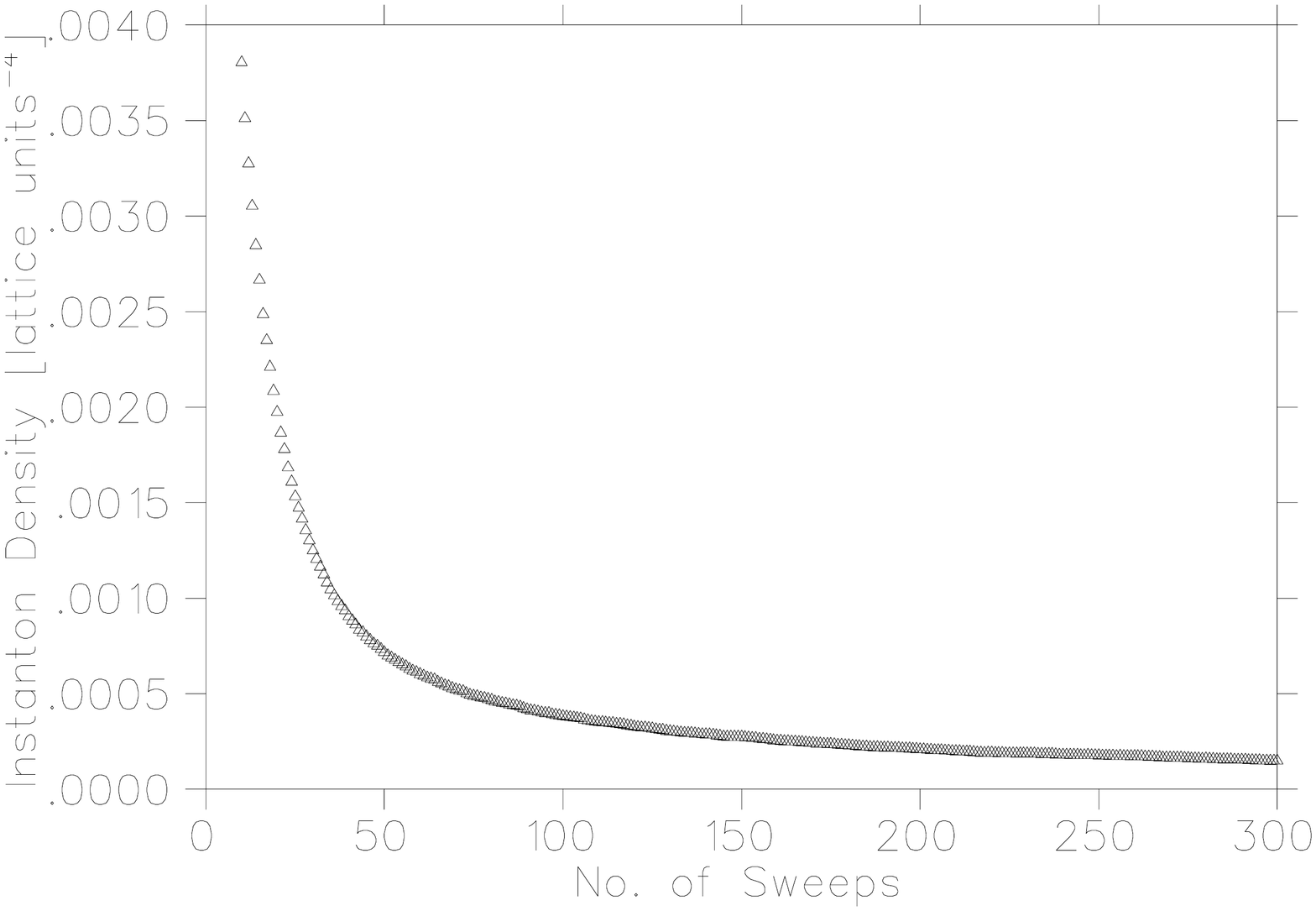}
\caption{The distance of instanton candidates from the theoretical relationship between charge at the centre and radius (left) and density of instanton candidates (right).}
\end{figure}

As illustrated in Fig.~\ref{fig:radiusvcharge}, after just 10 sweeps of smearing correlation with the theoretical charge/radius relationship is low. However, after 100 sweeps of smearing, all instanton candidates closely fit the theoretical lines. This gives us confidence that smearing can create a vacuum solely composed of instanton-like objects. To see how this unfurls over the course of smearing, we plot the Distance of each candidate from the Theoretical Relationship (DfTR) as well as the density of instanton candidates in Fig.~\ref{fig:tensweeps}.
\par
This shows an initial, rapid, decline in the DfTR and instanton candidate density, attributable to the loss of 'false positives'; local maxima of the action due to UV noise. This is also due to the loss of topologically non-trivial objects smaller than the dislocation threshold of $1$. After around 50 sweeps, we reach a point where we can be confident the gauge field configuration is dominated by instanton and anti-instanton like objects. We must, however, be wary of smearing eventually distorting the long-range structure of the vacuum.
\begin{figure}
\centering
\includegraphics[scale=0.275,angle=90]{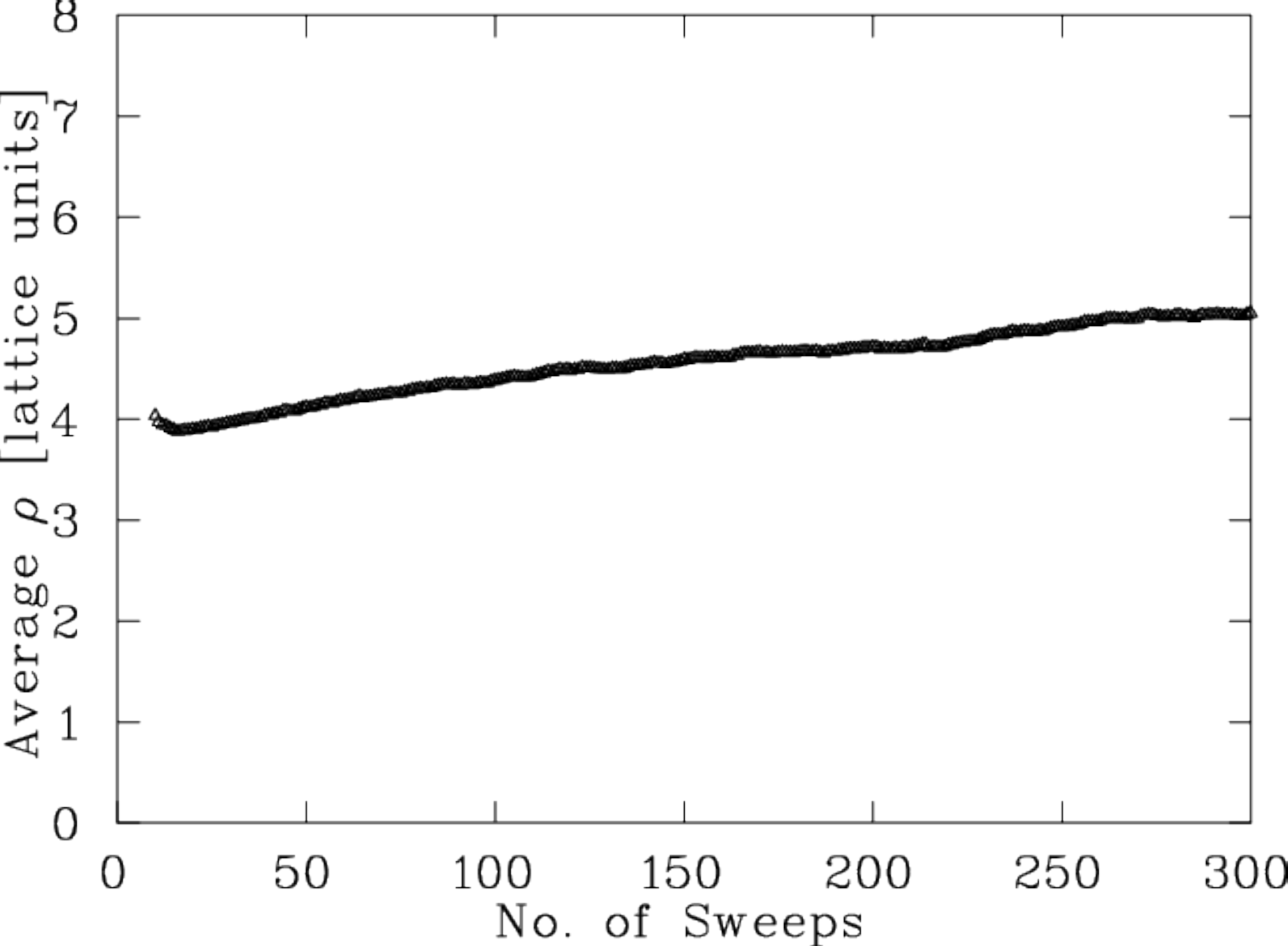}
\caption{Average radius of instanton candidates}
\label{fig:radius}
\end{figure}
\par 
We have plotted the average radius of instanton candidates in Fig.~\ref{fig:radius}, and can clearly see the enlarging effect of smearing occurring. An initial, rapid decline is most likely the result of false positives being removed, after which a steady increase is observed. We will choose to restrict our calculations to below 100 sweeps in order to minimise this effect, and choose 0, 30, 50, 80 and 100 sweeps to calculate the quark propagator.

\section{Results}
\label{sec:Results}
We use the overlap fermion action for the valence quarks, which has a lattice-deformed version of chiral symmetry, and hence no additive mass renormalisation \cite{Narayanan:1994}
\begin{equation}
D_{overlap}(\mu)=\frac{1}{2}(1-\mu)(1+\gamma_{5}\epsilon(\gamma_{5}D))+\mu.
\end{equation}
We use the FLIC action \cite{Zanotti:2002} as the overlap kernel, $D$, and present results for two values of $\mu$, $0.01271$ and $0.01694$, corresponding to input bare light quark masses of $0.0398$ GeV and $0.0530$ GeV. Simulations were performed on $50$ dynamical FLIC $20^{3}\times40$ configurations, with a lattice spacing of $0.126$ fm, corresponding to a spatial extent of $2.52$ fm. \newline
The overlap action provides a simple relationship between the quark propagator and the non-perturbative mass function, $M(p)$, defined by;
\begin{equation}
S(p) = \frac{Z(p)}{iq\!\!\!\!/\ + M(p)},
\end{equation}
containing all renormalisation information in $Z(p)$.

\begin{figure}
\includegraphics[scale=\scale,angle=90]{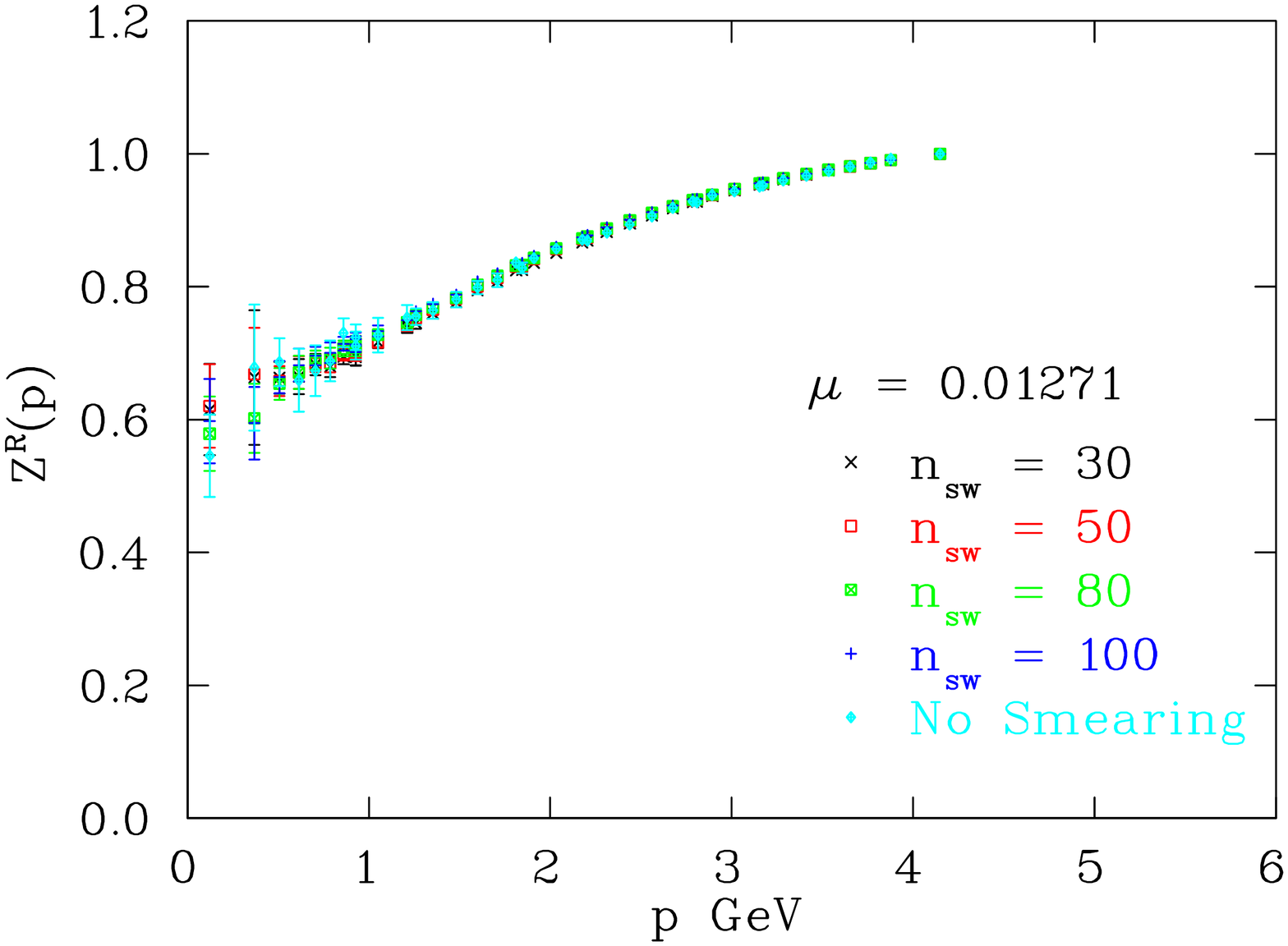}
\includegraphics[scale=\scale,angle=90]{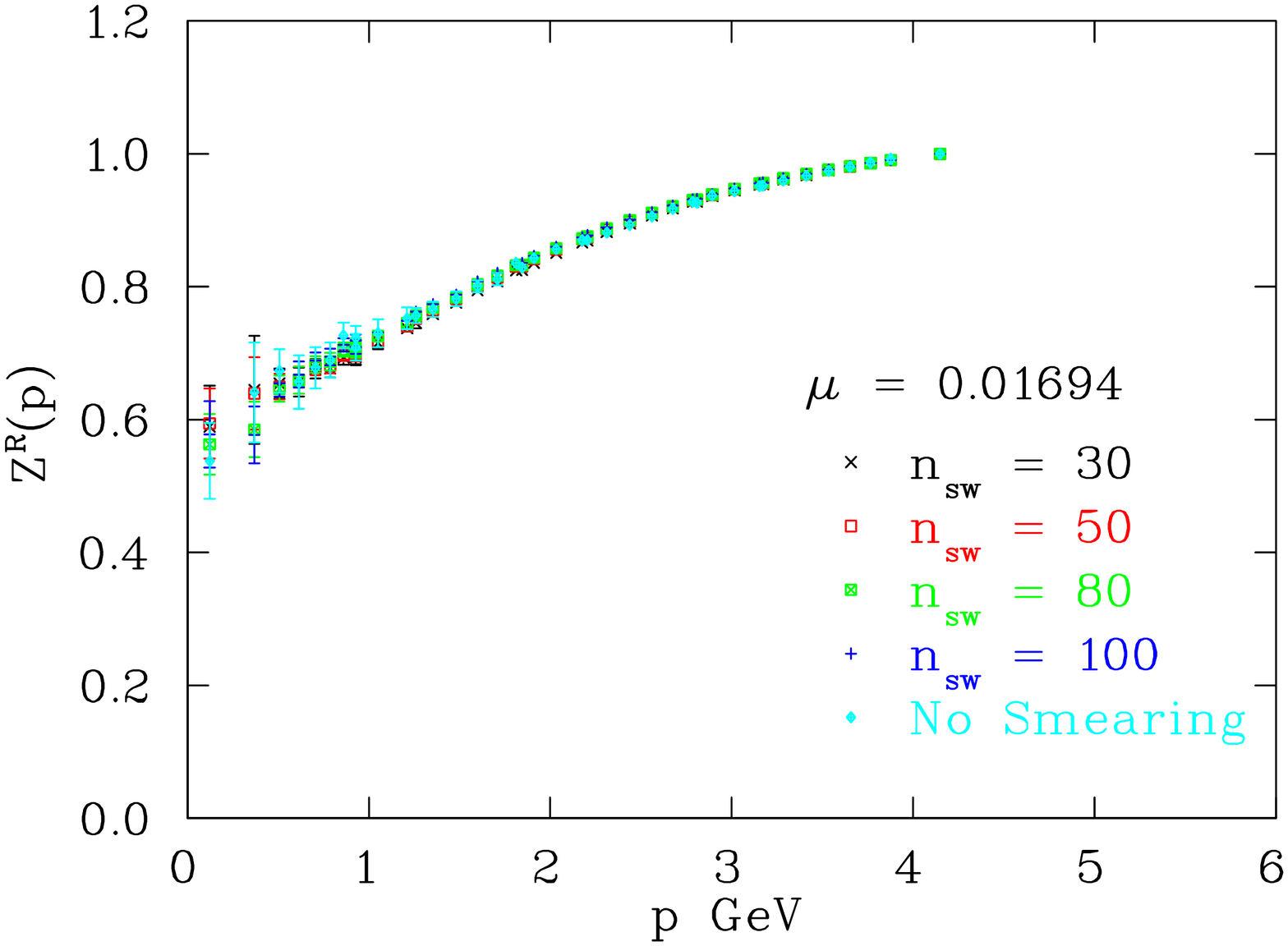}
\caption{Quark propagator renormalisation function at the two lowest masses considered, renormalised to 1 at the highest momentum point plotted.}
\end{figure}
\par

We have reproduced the well-known shape of the renormalisation function, showing little to no dependence on level of smearing.
\begin{figure}
\includegraphics[scale=\scale,angle=90]{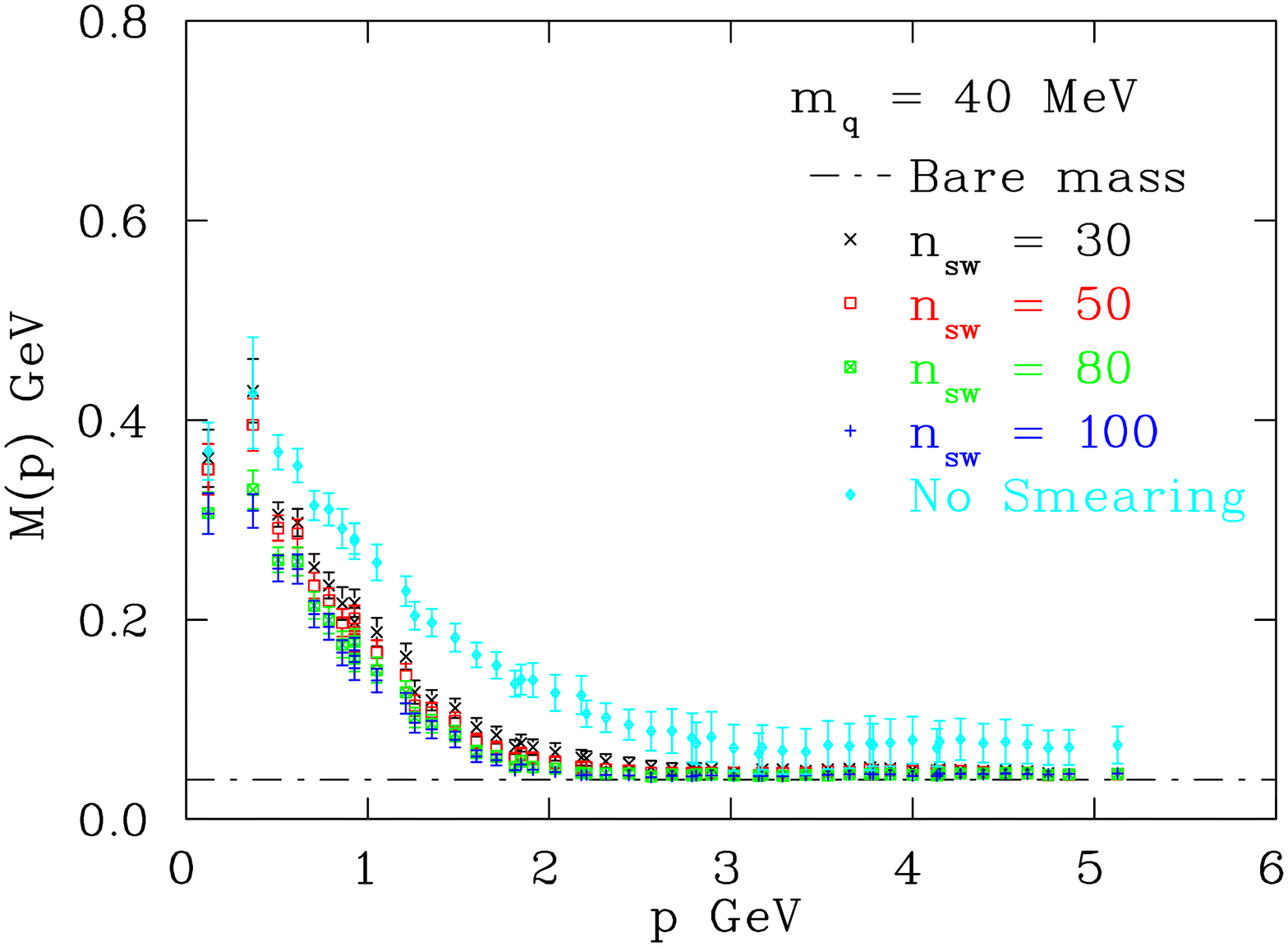}
\includegraphics[scale=\scale,angle=90]{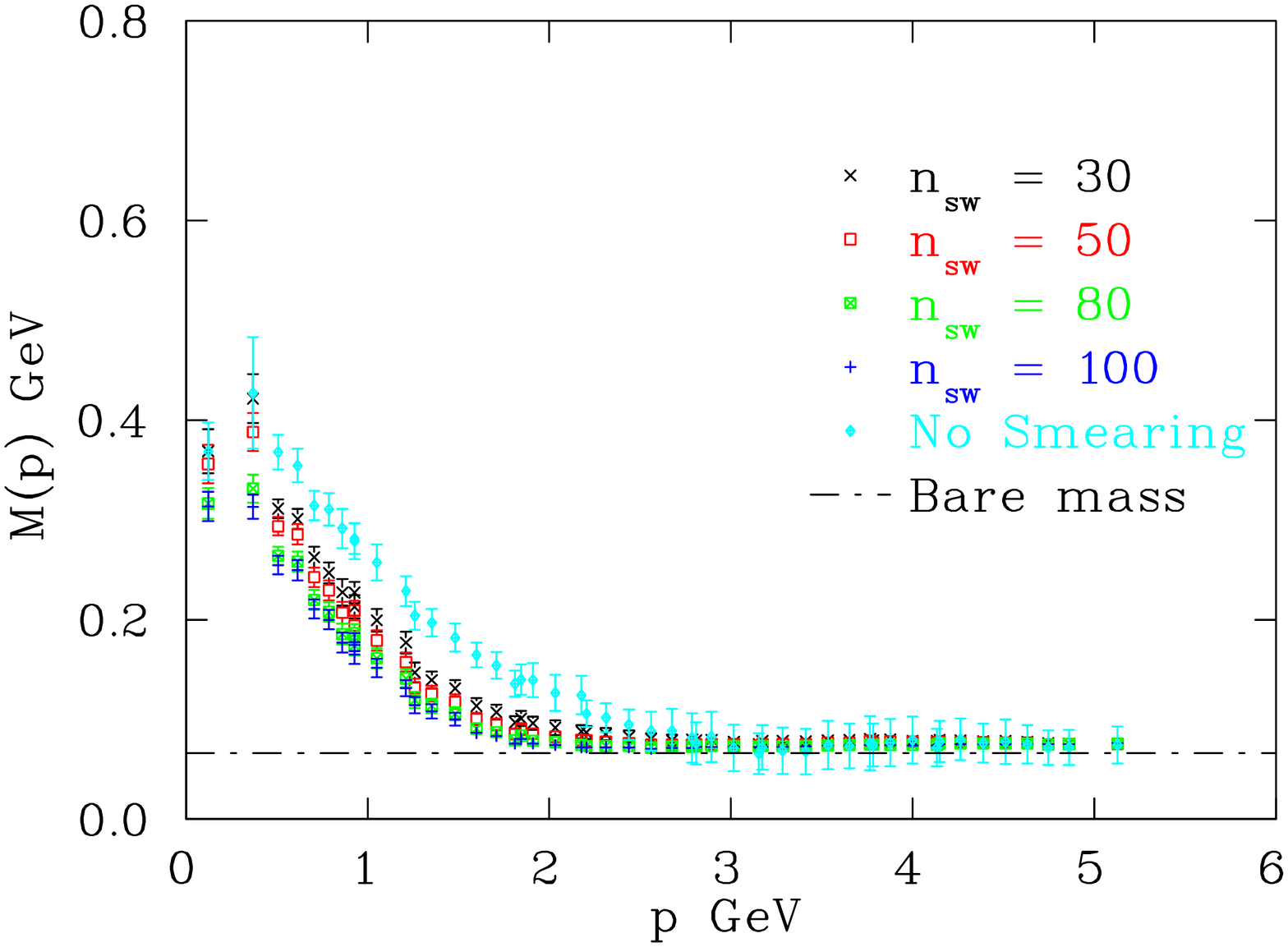}
\caption{Mass function at $\mu=0.01271$ (left) and shifted to match at high momenta (right).}
\end{figure}
\par
At high momenta, the quark has a running mass higher than the input bare mass due to short range gluon field effects. Smearing the lattice removes these, and the quark running mass on smeared configurations becomes equal to the input bare mass. We can see this effect clearly at $\mu=0.01271$; even after just 30 sweeps, on smeared configurations the mass function flattens to the input bare mass very rapidly. To facilitate comparisons at the low momenta we are interested in, we thus plot the smeared results for a value of $\mu = 0.02119$, corresponding to an input bare mass of $0.0662$ GeV, chosen to ensure agreement at high momenta. We can see that the majority of dynamically generated mass remains present on smeared configurations, even for high levels of smearing. This suggests that instantons are responsible for dynamical mass generation. There is however some loss, particularly at high levels of smearing. We attribute this to some removal of instanton degrees of freedom by the smearing algorithm, notably through instanton/anti-instanton annihilation. \newline
Portions of this work were carried out using eResearch SA and NCI National Facility computing resources. This research was supported by the ARC.


\begin{thebibliography} {99}

\bibitem{Berg:1981nw} 
  B.~Berg,
  Phys.\ Lett.\ B {\bf 104}, 475 (1981).

\bibitem{Teper:1985rb} 
  M.~Teper,
  Phys.\ Lett.\ B {\bf 162}, 357 (1985).
  
\bibitem{Ilgenfritz:1985dz} 
  E.~-M.~Ilgenfritz, M.~L.~Laursen, G.~Schierholz, M.~Muller-Preussker and H.~Schiller,
  Nucl.\ Phys.\ B {\bf 268}, 693 (1986).
  
\bibitem{Albanese:1987ds} 
  M.~Albanese {\it et al.}  [APE Collaboration],
  Phys.\ Lett.\ B {\bf 192}, 163 (1987).
  
\bibitem{Hasenfratz:2001hp} 
  A.~Hasenfratz and F.~Knechtli,
  Phys.\ Rev.\ D {\bf 64}, 034504 (2001)
  [hep-lat/0103029].
  
\bibitem{Morningstar:2003gk} 
  C.~Morningstar and M.~J.~Peardon,
  Phys.\ Rev.\ D {\bf 69}, 054501 (2004)
  [hep-lat/0311018].
  
\bibitem{Garcia Perez:1993ki} 
  M.~Garcia Perez, A.~Gonzalez-Arroyo, J.~R.~Snippe and P.~van Baal,
  Nucl.\ Phys.\ B {\bf 413}, 535 (1994)
  [hep-lat/9309009].
  
\bibitem{Moran:2008ra} 
  P.~J.~Moran and D.~B.~Leinweber,
  Phys.\ Rev.\ D {\bf 77}, 094501 (2008)
  [arXiv:0801.1165 [hep-lat]].
  
\bibitem{Garcia Perez:1993ki} 
  M.~Garcia Perez, A.~Gonzalez-Arroyo, J.~R.~Snippe and P.~van Baal,
  Nucl.\ Phys.\ B {\bf 413}, 535 (1994)
  [hep-lat/9309009].

\bibitem{Atiyah:1978physlett}
	M.F. Atiyah, N.J. Hitchin, V.G. Drinfield and Y.I. Manin,
	Phys. Lett. A {\bf 65}, 185 (1978)
	
\bibitem{Narayanan:1994} 
  R.~Narayanan and H.~Neuberger,
  Nucl.\ Phys.\ B {\bf 443}, 305 (1995)
  [hep-th/9411108].
	
\bibitem{Zanotti:2002} 
  J.~M.~Zanotti {\it et al.}  [CSSM Lattice Collaboration],
  Phys.\ Rev.\ D {\bf 65}, 074507 (2002)
  [hep-lat/0110216].
	
\end{thebibliography}
\end{document}